\documentstyle[preprint,aps]{revtex}
\begin{document}
\input psbox
\draft
\title{Fundamental Oscillation Periods of the Interlayer Exchange
Coupling beyond the RKKY Approximation}
\author{M. S. Ferreira \dag, J. d'Albuquerque e Castro\thanks
{Permanent address: Instituto de F\'{\i}sica, Universidade Federal 
Fluminense, Niter\'oi, Rio de Janeiro, 24001-970, 
Brazil}\ddag\hspace{0.1cm}, D. M. Edwards\dag \hspace{0.15cm} and 
J. Mathon \ddag }
\address{\dag  Department of Mathematics, Imperial College, London, 
SW7 2BZ, UK 
\\
\ddag City University, London, EC1V 0HB, UK }
\date{\today}
\maketitle
\begin{abstract}
A general method for obtaining the oscillation periods of the
interlayer exchange coupling is presented. It is shown that it is
possible for the coupling to oscillate with additional periods beyond
the ones predicted by the RKKY theory. The relation between the
oscillation periods and the spacer Fermi surface is clarified, showing
that non-RKKY periods do not bear a direct correspondence with the Fermi
surface. The interesting case of a FCC(110) structure is investigated,
unmistakably proving the existence and relevance of non-RKKY
oscillations. The general conditions for the occurrence of non-RKKY
oscillations are also presented.
\end{abstract}

\vspace{1cm}

\section{Introduction}
\label{introduction}

Oscillatory exchange coupling between metallic magnetic layers across a
non-magnetic spacer has been intensively studied over the last few years.
One of the main issues in this field has been the determination of the
oscillation periods of the coupling as a function of the spacer
thickness. The physical mechanisms proposed for explaining the
so-called oscillatory interlayer coupling include the quantum well
theory (QWT) of Edwards {\it et al.}\cite{ed1,ed2,mt1} and an extension
of the Ruderman-Kittel-Kasuya-Yosida (RKKY) theory adapted to the
multilayer geometry due to Bruno and 
Chappert\cite{bruno1,bruno2}. 

Both theoretical approaches seem to agree that the origin of
the oscillation periods is in the structure of the spacer Fermi
surface (FS). In fact, in the RKKY theory the
periods are associated with wave vectors directly obtained from the FS.
More specifically, they are given by vectors perpendicular to the layer
planes which span two points of the FS with anti-parallel Fermi
velocities\cite{bruno1,bruno2}. There has been a general belief that all
oscillation periods are given by the RKKY theory and therefore that such
a simple picture is always valid. Indeed, it has been shown that in
certain simple models \cite{ed1,ed2,mt1} the periods predicted by the
QWT coincide with the RKKY ones and are given by the extremal dimensions
of the spacer FS in the direction perpendicular to the layers. In
another case \cite{fcc111}, where the lattice lacks reflection symmetry
about the layer planes, the correspondence between the quantum well and
the RKKY periods is rather subtle but still obtains. In the models
mentioned above, harmonics of the RKKY periods are obtained but no
new fundamental periods. Furthermore, d'Albuquerque e Castro {\it et
al}. \cite{paper1} showed analytically for a very general model that
RKKY theory holds in the limit of small exchange splitting in the
ferromagnetic material. 

However, using a more general approach\cite{paper1,torque} based on the
existence of quantum well states\cite{ed1,ed2,mt1}, fundamental periods
not predicted by RKKY theory have been recently 
obtained for the particular case of a FCC lattice along the (110)
direction \cite{paper3}. In that communication, within the single-band
tight-binding model with nearest neighbour interaction, it is shown 
that in addition to the RKKY predictions, a number of distinct periods is
involved in the coupling. Those new periods arise from non-negligible
higher order terms which RKKY, as a second order perturbative approach,
does not account for.
The resemblance of the FS considered with those of the noble metals
suggests possible real effects in systems having these materials as
spacers. Moreover, the occurrence of non-RKKY behaviour within
the simplest single-band model indicates that it should also
happen in the more elaborate multi-band case. In fact, non-magnetic
transition metals with their multi-sheet Fermi surfaces are suitable
candidates for presenting non-RKKY periods. Furthermore, in most cases
investigated so far, each oscillation period comes from distinct ${\bf
q}_\parallel$-points of the 2-dimensional Brillouin zone (BZ). Even when 
multi-periodic oscillations occur, one value of ${\bf q}_\parallel$
yields only one period. The existence of more than one period for a
single value of ${\bf q}_\parallel$, which may arise in a number of
cases, has not been thoroughly investigated and the combination of such
periods may lead to new features of the oscillations. For the reasons
above, a general understanding of the mechanisms determining all
oscillation periods is needed. 

The purpose of this work is to generalize the method previously used to
calculate the oscillation periods of the interlayer coupling. The
relation between the periods and the spacer FS is clarified and in
addition to the determination of the usual periods, the general
conditions for occurrence of non-RKKY behaviour are also established.
Furthermore, here we present a rather detailed set of results in order
to illustrate better the different stages of the calculations. The
sequence of the paper is as follows. In section \ref{general} we present
a general method for obtaining the oscillation periods of the interlayer
coupling, where in addition to the determination of the usual periods,
the existence of non-RKKY behaviour is proved to be possible. Similarly
to previous theoretical approaches\cite{bruno1,bruno2,ed1,mt1,spa2}, an
expression for the coupling in the asymptotic limit of large spacer
thickness is derived. Here however, the occurrence of multi-periodicity
for a single value of ${\bf q}_\parallel$ becomes evident at the same
time that the relation between the oscillation periods and the
spacer FS is clarified. In section \ref{fcc110}, in order to
illustrate the method, we look at the particular case of the FCC lattice
along the (110) direction which unmistakably shows the relevance of
terms beyond RKKY approximation. Finally, in the
last section we conclude and present the general conditions for
occurrence of non-RKKY oscillations. 

\section{General method}
\label{general}

Based on the formalism introduced by d'Albuquerque e Castro {\it et
al.}\cite{paper1}, where the coupling is expressed in terms of the
one-electron Green's functions of the multilayered system, we here
present a general method for determining all the oscillation
periods involved in the coupling. 

For the sake of simplicity, we consider a system consisting of two
parallel magnetic planes embedded in an infinite non-magnetic material.
We label the two ferromagnetic planes $0$ and $n$, so that the number
of non-magnetic spacer planes is $n - 1$. As far as the coupling as a
function of the spacer thickness is concerned the number of magnetic
planes does not affect the periods \cite{zeetal}, influencing only phase
and amplitude of the oscillations. Since the periods are our main
concern, the restriction to two single magnetic planes does not pose any
limitations on the results here obtained.

The expression for the coupling $\Delta \Omega(\pi)$,
namely the difference in the thermodynamical potential $\Omega$
between the ferromagnetic and antiferromagnetic configurations, is given
by\cite{paper1} 
\begin{equation}
\Delta \Omega(\pi) \,=\, {1\over\pi} \,\sum_{{\bf q}_\parallel} \,
\int d\,\omega\,f(\omega)\,{\cal F}({\bf q}_\parallel,\omega)\,\,\,,
\label{j}
\end{equation}
where the function ${\cal F}$ is given by 
\begin{equation}
{\cal F}({\bf q}_\parallel,\omega) = Im\, tr \, \ln\,\{\,\, 1 + 4 
\,G^{\uparrow}_{n 0}({\bf q}_\parallel,\omega) \, V_{ex} \, 
G^{\downarrow}_{0 n}({\bf q}_\parallel,\omega) \, V_{ex}\}\,.
\label{Fwq}
\end{equation}
In the equations above $G^{\sigma}_{n 0}({\bf q}_\parallel,\omega)$ is
the propagator between planes $0$ and $n$ for electrons with spin
$\sigma$ in the ferromagnetic configuration of the system, $f(\omega)$
is the Fermi function, $V_{ex}$ is a diagonal matrix in orbital indices
representing the strength of the local exchange potential in the
ferromagnetic layers, and the summation over ${\bf q}_\parallel$ is
restricted to the two-dimensional Brillouin zone (BZ). It is worth
stressing that all matrices above are in orbital indices. We assume for
simplicity that the non-magnetic part of the potential in the
ferromagnetic planes is identical to the spacer on-site potential. This
is not an essential assumption and can be lifted without major effects
on the results. In fact, it has been shown that the presence of a
spin-independent square well does not influence the periods of the
interlayer coupling, affecting only the phase and amplitude of the
oscillation \cite{paper2}. As such, the exchange splitting of the
magnetic material acts as two localized perturbations on the planes $0$
and $n$ and in this situation the propagators $G$ can be written as a
function of the bulk spacer propagators $g_{0 n}$ and $g_{n 0}$, {\it
i.e.}, 
\begin{equation}
G^{\sigma}_{n 0}({\bf q}_\parallel,\omega) = G^\sigma \left( g_{0
n}({\bf q}_\parallel,\omega), g_{n 0}({\bf q}_\parallel,\omega) \right) .
\label{Ggeneral}
\end{equation}
Within the particular case of the single-band model the explicit
dependence of $G^\sigma$ upon the off-diagonal propagators $g_{0 n}$ and 
$g_{n 0}$ is, for each ${\bf q}_\parallel$ and $\omega$, given by
\begin{equation}
G^{\uparrow ,\downarrow}_{n 0} ={\pm}{\tau_{\uparrow,\downarrow}\,
g_{n 0}\,(1-\tau_{\uparrow,\downarrow}\,g_{0 0}) \over
V_{ex}\,(1-\tau_{\uparrow,\downarrow}\,g_{n 0} \tau_{\uparrow,\downarrow}
\,g_{0 n})} \,,
\label{G}
\end{equation}
where $g_{\ell m}$ is a general matrix element of the bulk spacer
Green's function and $\tau_{\uparrow,\downarrow} = {\pm} V_{ex}(1 {\pm}
V_{ex}g_{0 0})^{-1}$. An analogous expression can be obtained for
$G^{\uparrow,\downarrow}_{0 n}$. Note that whereas Eq.(\ref{Ggeneral})
involves matrices in orbital indices, all terms in Eq.(\ref{G}) are
scalars. 

Returning to the general case and for reasons of future comparison, it is
worth presenting the RKKY-limit expression for the bilinear coupling
$J_1^{RKKY}$\cite{paper1,paper2}. It is given by Eq.(\ref{j}) with
${\cal F}$ replaced by
\begin{equation}
{\cal F}_{RKKY}({\bf q}_\parallel,\omega) = Im \,\, tr \,
\left\{\,2\,V_{ex}^2\, g_{n 0}({\bf q}_\parallel,\omega) \, g_{0 n}({\bf
q}_\parallel,\omega) \right\}\,.
\label{Frkky}
\end{equation}
We recall that the expansion of $\Delta \Omega(\theta)$ in powers of
$\cos(\theta)$, being $\theta$ the angle between the magnetization
vectors in the magnetic planes, defines the bilinear and biquadratic
couplings $J_1$ and $J_2$, respectively \cite{heinrich}. Moreover, it is
found that higher order terms than $\cos^2(\theta)$ are negligible so
that $J_1 = \Delta \Omega(\pi)/2$ \cite{paper2}. 

It is clear from Eqs.(\ref{Fwq}), (\ref{Ggeneral}) and (\ref{Frkky})
that in either case the spacer thickness dependence of the interlayer
coupling is entirely determined by the off-diagonal propagators $g_{0
n}$ and $g_{n 0}$. The investigation of the oscillation periods thus
requires an analysis of $g_{n 0}$ as a function of $n$. 

\subsection{Calculation of the 1-dimensional propagators}

For fixed values of ${\bf q}_\parallel$, the Green's functions
$g$ are equivalent to 1-dimensional propagators with ${\bf
q}_\parallel$-dependent hoppings and in-plane energies. 
We recall that
$g$ is a matrix over orbital indices and therefore a general matrix
element $g_{\ell m}^{\mu \nu}$ represents the propagation of an electron
with orbital $\mu$ from plane $\ell$ into an orbital $\nu$ at plane $m$.
It is given by
\begin{equation}
g_{\ell m}^{\mu \nu}({\bf q}_\parallel,\omega) = ({d \over 2\pi}) \,
\sum_s \, \int_{-{\pi \over d}}^{\pi \over d}\,\,dq_\perp \, {a_{s
\mu}^*(q_\perp) \, a_{s \nu}(q_\perp) \, e^{-i q_\perp (\ell-m) d}
\over \omega - E_s({\bf q}_\parallel,q_\perp) + i\,0^+}\,\,,
\label{integral}
\end{equation}
where $q_{\perp}$ is the wave vector perpendicular to the layers, $d$ is
the interplane distance, $E_s({\bf q}_\parallel,q_\perp)$ describes
the bulk spacer band structure, being $s$ the band index, and $a_{s
\mu}(q_\perp) \equiv \langle q_\perp \, s \, \vert q_\perp \, \mu
\rangle$ is the projection of the eigenstate $s$ into the orbital $\mu$
for a given $q_\perp$.  

The integral in Eq.(\ref{integral}) can be evaluated for $\ell < m$ by
simply extending $q_\perp$ to the complex plane and changing the 
contour integration from a straight line on the real axis to the
boundaries of a semi-infinite rectangle in the upper-half plane whose
base lies on the real axis between $-\pi/d$ and $\pi/d$. For the case
$\ell > m$, the extension is to a rectangle in the lower-half plane. In
either case the integrand vanishes as 
$\vert Im \,q_\perp \vert \rightarrow \infty$ and because $q_\perp =
-{\pi \over d} + i \, y$ 
and $q_\perp = {\pi \over d} + i \, y$ are equivalent wave vectors, the
integrals along the vertical sides of the rectangle cancel each other,
simplifying the problem to the calculation of the residues associated
with the poles of the integrand inside the closed contour. The poles are
in turn given by the values of $q_\perp$ satisfying the condition
$\omega^+ - E_s({\bf q}_\parallel,q_\perp) = 0$, where $\omega^+ =
\omega + i \, 0^+$. 

The poles are clearly dependent on the band structure. Due to the
in-plane symmetry of the system, the general tight-binding Hamiltonian
of the homogeneous system is written as 
\begin{equation}
H({\bf q}_\parallel,q_\perp) = \sum_{r r^\prime} \, H_{r r^\prime}({\bf
q}_\parallel) \, e^{i \, q_\perp \,( r - r^\prime )\, d} \,\,,
\label{hamiltonian} 
\end{equation}
where $H_{r r^\prime}({\bf q}_\parallel)$ are operators describing the
electron hopping between planes $r$ and $r^\prime$. The upper limit of
the sum depends on the maximum number of planes ${\bar r}$ connected
through the electron hoppings as well as on the lattice structure. As we
shall see, in the FCC lattice along the (110) direction for instance,
even with only hoppings between nearest atoms being included, electrons
may hop between nearest and next nearest planes. For a fixed value of
$\omega^+$, the wave vectors $q_\perp$ satisfying the equation $\omega^+
- E_s({\bf q}_\parallel,q_\perp) = 0$ are the solutions of a polynomial
equation in $e^{i \, q_\perp \, d}$, {\it i.e.}, 
\begin{equation}
\sum_{r}^{\bar r} \, \Upsilon_r({\bf q}_\parallel) \,\, e^{i \, q_\perp
\, r \, d} = 0 \,\, ,
\label{w+-E} 
\end{equation}
where $\Upsilon_r({\bf q}_\parallel)$ is a function of the ${\bf
q}_\parallel$-dependent hoppings and in-plane energies. Note that the
solutions of the equation above are exponentials $e^{i \, q_\perp \, d}$
and not wave vectors $q_\perp$ themselves. We label the poles $q_j^s$,
where $j$ numbers the poles of a given band index $s$. Since
$E_s(q_\perp)=E_s(-q_\perp)$ poles appear in pairs and may be real or
complex. We see from a simple Taylor expansion of $\omega - E_s({\bf
q}_\parallel,q_\perp)$ around the real poles that, for a given pair,
which pole contributes to the integral depends on the sign of the
derivative $(\partial E_s/\partial q_\perp)_{q_j^s}$. For a given band
$s$ which crosses the energy $w$ twice, yielding two pairs of real
poles, the contributory poles $q_1^s$ and $q_2^s$ have different signs. 
In other words, when $q_1^s$ is in the range [$0,\pi$], $q_2^s$ is in
the range [$-\pi,0$] or vice versa. 

It is interesting to look at the physical significance of the real poles.
Being obtained from the band structure $E_s({\bf q}_\parallel,q_\perp)$,
the poles $q_j^s$ are just the wave vectors with which electrons of
energy $\omega$ having in-plane energies $\epsilon({\bf q}_\parallel)$
and hoppings $t({\bf q}_\parallel)$ propagate across the spacer. Thus,
for $\omega = E_F$, where $E_F$ is the Fermi energy, the real poles are
just the perpendicular coordinates of the FS for a fixed value of ${\bf
q}_\parallel$. As discussed later, complex poles contribute
significantly only in the case of thin spacers.

By adding the residues associated with the contributory poles we obtain
an analytical expression for $g_{\ell m}$, given by
\begin{equation}
g_{\ell m}({\bf q}_\parallel,\omega) = \sum_s \, \sum_{j} A_{s j} \,
e^{-i q_j^s (\ell-m) d}\,\,\,,
\label{glm}
\end{equation}
where the matrix elements of the matrix $A_{s j}$ are
\begin{equation}
A_{s j}^{\mu \nu}({\bf q}_\parallel,\omega) =  \,-\,i\,d\,\,a_{s
\mu}^*(q_j^s)\,a_{s \nu}(q_j^s)\left\{\left[\,{\partial E_s({\bf
q}_\parallel,q_\perp) \over \partial q_\perp}\,\right]
_{q_\perp=q_j^s}\right\}^{-1}\,\,.
\label{AA} 
\end{equation}
The coefficients $A_{s j}^{\mu \nu}$ depend neither on $\ell$ nor on $m$
and the only dependence on the distance between the planes is inside the
argument of the exponentials. In addition, the exponentials are
independent of the orbital indices, which means that all matrix elements
oscillate with the same periods. Note that in calculating Eq.(\ref{AA})
we have assumed that the coefficients $a_{s \mu}(q_\perp)$ are not
ill-behaved, that is, they have neither singularities nor branch points
inside the integration contour. In this way, we just need to evaluate
the coefficients at the poles $q_j^s$.

\begin{figure}
\begin{center}\mbox{\psboxscaled{800}{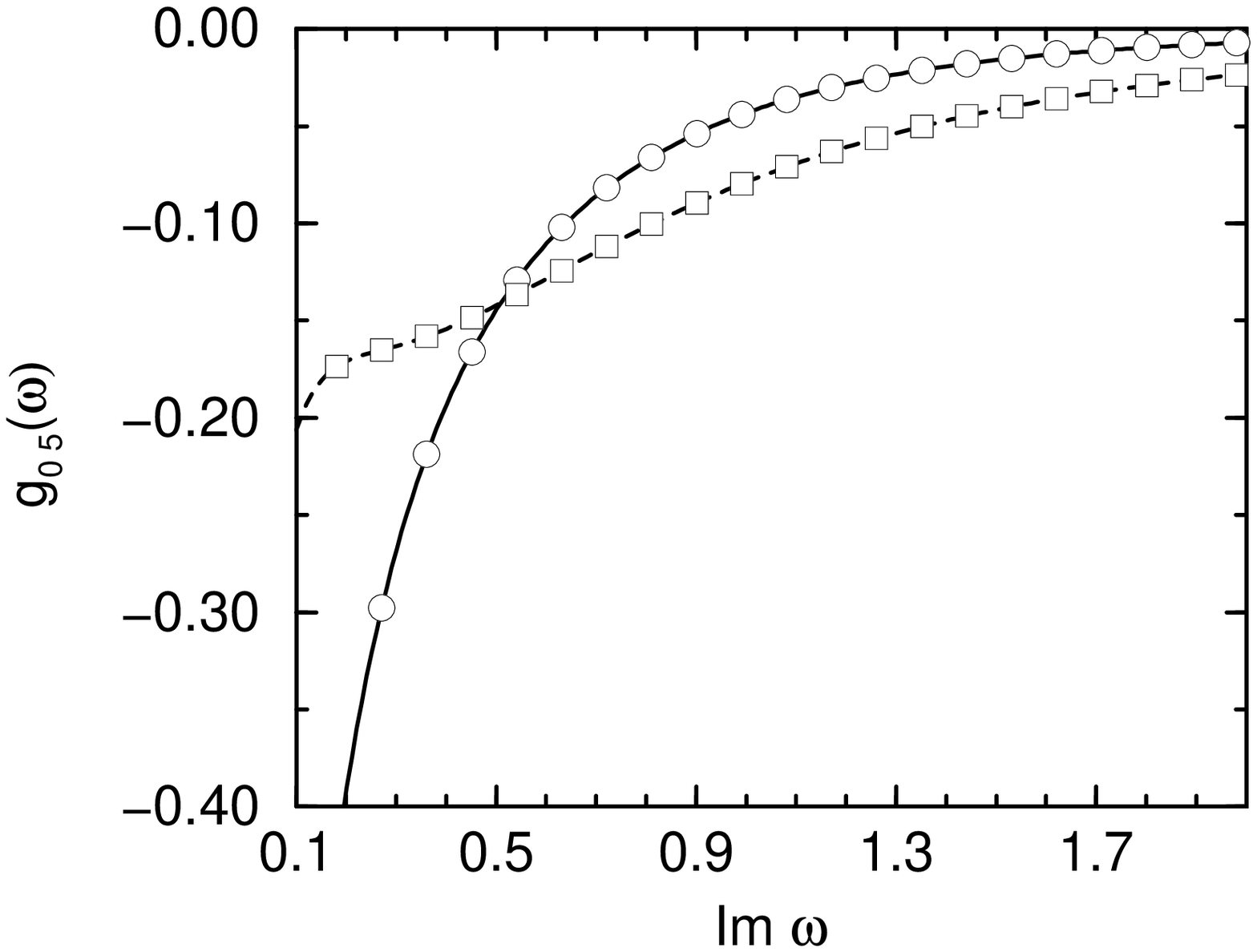}}\end{center}
\caption{Comparison between the analytical expression and the
brute-force calculation of the propagator $g_{0 5}(\omega)$. The lines
represent the analytical results and the symbols the numerical
ones. The full line and the circles are the real part of $g_{0 5}$
whereas the dashed line and the squares are the imaginary part. The
parameters of this tight-binding Hamiltonian are indicated in the text.
In order to improve the convergence of the brute-force calculations we
have varied $\omega$ along the imaginary axis.} 
\label{g05}
\end{figure}

Nevertheless, in a general case the coefficients $a_{s \mu}(q_\perp)$
are not well behaved. Although they do not have singularities like in
the denominator of Eq.(\ref{integral}), branch points and
cuts do exist. The contributions from the branch points
must be taken into account otherwise a simple summation of the residues
will not give the correct result of the integration. This is a very
important point that, if overlooked, may lead to erroneous results. In
fact, in a paper by Bruno \cite{brunoso'} where the coupling is
expressed in terms of transmission and reflection coefficients of
electrons across quantum barriers, a similar integral to that in
Eq.(\ref{integral}) arises. In that communication no particular
attention was given to the analytic properties of the coefficients $a_{s 
\mu}(q_\perp)$ and the result above was erroneously concluded to be
exact for arbitrary values of $\ell$ and $m$. It is important to stress
that Eqs.(\ref{glm}) and (\ref{AA}) are not correct in general. The
equations above become exact for all $\ell$ and $m$ only within the
single-band model, where there are no coefficients introducing branch
points. This is clearly seen in Fig.\ref{g05} where the real and
imaginary parts of the function $g_{0 5}(\omega)$ of a linear chain with
hoppings up to four nearest neighbours within the single-band model are
displayed. The parameters used were $\epsilon=0$, $t_1=-1/2$,
$t_2=-1/2$, $t_3=1/5$ and $t_4=1/10$, where $\epsilon$ is the in-plane
energy and $t_\ell$ are the hoppings to the $\ell$-th neighbour planes.
The analytical results following Eq.(\ref{glm}), represented by the
lines, are in absolute agreement with the brute-force calculation of
Eq.(\ref{integral}), represented by the symbols. In the general case
there are also contributions to the $q_\perp$ integral arising from
branch cuts associated with the coefficients $a_{s \mu}(q_\perp)$. Since
in general these cuts do not intersect the real axis, the factor $e^{i
\, q_\perp ( \ell - m ) \, d}$ in the integrand of Eq.(\ref{integral})
ensures that the contribution is negligible for large $\vert \ell - m
\vert$. Thus, failure to treat the branch cuts correctly does not
introduce errors in the asymptotic limit of large $\vert \ell - m
\vert$.  

It is clear that $g_{\ell m}$ is an oscillatory
function of the distance between the planes and it oscillates with
different periods, each one associated with its respective wave vector
$q_j^s$. In addition, the combination of exponentials allows a
quasi-periodic behaviour of $g_{\ell m}$ because, in general, the wave
vectors $q_j^s$ are incommensurate. This fact, as we shall see, may
have a striking effect on the oscillatory coupling. Another point to be
mentioned is the possibility of evaluating $g_{\ell m}$ in
Eq.(\ref{glm}) for non-integer values of $\ell$ and $m$, which allows
the evaluation of the coupling for continuous values of spacer
thickness.

We now mention the case of complex poles. Since they have non-zero
imaginary parts, their contribution for large separation between the
planes is again negligible. Thus only real wave vectors contribute to the
oscillations for large $\vert \ell - m \vert$. Bearing in mind that the
principal contribution to the coupling comes from the energy region
around the Fermi level\cite{ed1,ed2,mt1,spa2}, it is clear that at least
one real wave vector will be involved in the expression for $g_{\ell
m}$. Thus the sum over the band index $s$ in Eq.(\ref{glm}) must be
only over the bands which cross the Fermi level. 

\subsection{Multiple Fourier expansion}

Having investigated the function $g_{\ell m}$ as a function of the distance
between the planes, we recall that ${\cal F}$ is expected to oscillate
with the same periods. The obvious way of representing the function
${\cal F}$ is through a simple Fourier expansion, which nevertheless
cannot be done due to the quasi-periodicity of $g$. In other words,
because the wave vectors $q_j^s$ are in general incommensurate, $g_{0
n}$ and $g_{n 0}$ may have an overall non-periodic behaviour. To
overcome this problem we make use of the multiple Fourier series,
recently used to deal with the problem of magnetic thickness dependence
of the interlayer coupling\cite{zeetal}. The procedure consists in
replacing the spacer thickness $n$ multiplying the wave vectors $q_j^s$
in each exponential with fictitious spacer thicknesses $n_j^s$, such
that Eq.(\ref{glm}) becomes  
\begin{equation}
\bar{g}(n_1^1,...,n_j^s,...) = \sum_s \, \sum_j A_{s j} \,
e^{-i \, q_j^s \, n_j^s \, d}\,\,\,.
\label{gbar}
\end{equation}
The new function $\bar{g}$ is now periodic in each variable $n_j^s$
separately, and so is its corresponding $\bar{{\cal F}}$. Bearing in
mind that $\bar{g}$ does not correspond to the physically real case,
which is reproduced by making $n_j^s=n$ for every $j$ and $s$, we
evaluate the Fourier series for each variable $n_j^s$ separately,
eventually making them all equal $n$. ${\cal F}$ then becomes 
\begin{equation}
{\cal F} = \sum_{(m_1^1,...,m_j^s,...)}\, C_{(m_1^1,...,m_j^s,...)} \,\,
e^{i \, n  \, d \, \sum_{s j} (m_j^s \, q_j^s)}\,\,,
\label{f}
\end{equation}
where $m_j^s$ are integers, $C_{(m_1^1,...,m_j^s,...)}$ are the
Fourier coefficients given by 
\begin{equation}
C_{(m_1^1,...,m_j^s,...)} = \left({1 \over \Lambda_1^1 ... \Lambda_j^s
...}\right)\,\int_0^{\Lambda_1^1} d n_1^1 ...\int_0^{\Lambda_j^s} d
n_j^s \,... \,\, \bar{{\cal F}}(n_1^1,...,n_j^s,...)\, e^{i \, d \sum_{s
j}(m_j^s \, q_j^s \, n_j^s)} \,\,,
\label{coefs}
\end{equation} 
and $\Lambda_j^s = 2 \pi/\vert q_j^s \vert$ are the periods associated
with the wave vectors $q_j^s$. The number of indices involved in the
Fourier coefficients is the same as the number of poles because each
$q_j^s$ has been associated with a fictitious spacer thickness $n_j^s$.
In spite of the somewhat congested notation, in practice there are only
a few indices in the coefficients due to the restricted number of real
solutions crossing the Fermi level. In fact, within the single-band
model the calculation of the Fourier coefficients can be extremely
simplified, as shown in appendix \ref{fourier}. The analysis of such
a simplified model, although primarily illustrative, reflects some
properties common to the more general case. Two of these properties are
$C_{-m_1,-m_2} = C_{m_1,m_2}^*$ and $C_{m_1,m_2} = 0$ for odd values of
$m_1 + m_2$. 

\subsubsection{RKKY-Coefficients}
\label{rkky}

We now focus on the RKKY-limit expression for the coupling. 
This is the limit of small ferromagnetic exchange splitting. 
It is clear from Eq.(\ref{Frkky}) that unlike the coupling $\Delta
\Omega$, the RKKY expression involves neither logarithmic functions 
nor combinations of $G_{n 0}^\sigma$ and $G_{0 n}^\sigma$. In fact,
${\cal F}_{RKKY}$ simply contains a single product of the propagators
$g_{n 0}$ and $g_{0 n}$. Bearing Eq.(\ref{Frkky}) in mind we define
${\cal F}_{RKKY} \equiv  [{\cal F}^\prime_{RKKY} - ({\cal
F}^\prime_{RKKY})^*]/2 i$ and from the expression for an arbitrary
element $g_{\ell m}$ in Eq.(\ref{glm}) we write
\begin{equation}
{\cal F}^\prime_{RKKY} = 2 \, V_{ex}^2 \sum_{\mu \nu} \, \sum_{s,
s^\prime} \, \sum_{j, j^\prime} A_{s j}^{\mu \nu} \, A_{s^\prime
j^\prime}^{\nu \mu} \,\, e^{i(q_j^s+q_{j^\prime}^{s^\prime})n d} \,\,. 
\label{fprime}
\end{equation}
Clearly, the above equation shows that the integrand ${\cal F}_{RKKY}$
can be manipulated and put in a Fourier expansion form, similarly to
${\cal F}$ in Eq.(\ref{f}). However, the main difference is that in this
case the number of terms involved in the expansion is limited, related
to the maximum number of poles in Eq.(\ref{glm}), whereas the 
sum in Eq.(\ref{f}) runs over an infinite number of terms. Such
difference is due to the fact that the RKKY-limit expression results 
from a perturbative approach corresponding to the lowest order term of
an expansion. All the terms included in ${\cal F}_{RKKY}$ are also
included in the full expression ${\cal F}$. The higher order terms
correspond to oscillations that are either harmonics of the fundamental
RKKY behaviour or effectively new oscillation periods. This point
confirms the RKKY theory as a limit of the more general approach used
here. We also point out that in this case, the argument of the 
exponentials has the form $(m_j^s q_j^s + m_{j^\prime}^{s^\prime}
q_{j^\prime}^{s^\prime})$, involving only two wave vectors and with the
constraint $\vert m_j^s + m_{j^\prime}^{s^\prime} \vert = 2$, being
$\vert m_j^s \vert \le 2$ and $\vert m_{j^\prime}^{s^\prime} \vert \le
2$. This is a consequence of ${\cal F}_{RKKY}$ being the product of only
two propagators, namely, $g_{n 0}$ and $g_{0 n}$. In the more general
case of ${\cal F}$, the infinite possibilities of combining the
mentioned propagators does not pose such a restriction. 

At this point we raise the question of whether the RKKY periods of the
interlayer coupling remain valid as the exchange splitting of the
ferromagnetic layers, associated with $V_{ex}$, is increased. The
question can be reformulated as whether the higher order terms in
${\cal F}$, apart from the RKKY harmonics, do contribute to the
coupling. The answer to this question is in the evaluation of the
integrals in Eq.(\ref{j}). 

\subsection{Stationary phase approximation}

Having written ${\cal F}$ (and ${\cal F}_{RKKY}$) in terms of a Fourier
expansion, we can now make use of the stationary phase
method\cite{ed1,ed2,mt1,spa2}, which yields an analytical expression for
the coupling, asymptotically exact for thick spacers ($n >> 1$).

Inserting Eq.(\ref{f}) into (\ref{j}), we have
\begin{equation}
\Delta \Omega(\pi) \,=\, {1\over\pi}\sum_{(m_1^1,...,m_j^s,...)} 
\,\sum_{{\bf q}_\parallel} \, \int d\,\omega\,f(\omega)\,
C_{(m_1^1,...,m_j^s,...)} \,\, e^{i \, n \, d \, \sum_{s j} (m_j^s \,
q_j^s)} \,\,. 
\label{statpa}
\end{equation}
We recall that the coefficients $C_{(m_1^1,...,m_j^s,...)}$ and the wave
vectors $q_j^s$ depend on both integration variables ${\bf q}_\parallel$
and $\omega$. The stationary phase method shows that in the asymptotic
limit of large spacer thickness the only contributions to the integral
over ${\bf q}_\parallel$ arise from the neihbourhood of points which
make the arguments of the exponentials stationary. We call these points
${\bf q}_\parallel^\alpha$, where the superscript $\alpha$ labels the
different vectors. Furthermore, the integral over $\omega$ is also
non-negligible only around the Fermi energy $E_F$. Hence, the wave
vectors that effectively contribute to the coupling must satisfy the
following condition,
\begin{equation}
\sum_{s j} \left[\,m_j^s \; {\bf \nabla}_\parallel \, q_j^s({\bf
q}_\parallel,E_F)\right] \, = 0 \,\,,
\label{grad}
\end{equation}
where ${\bf \nabla}_\parallel$ is the two-dimensional gradient in ${\bf
q}_\parallel$-space. In other words, the equation above gives the
necessary conditions for a constructive interference between the
electrons across the spacer. The otherwise destructive interference does 
not contribute to the coupling. Eq.(\ref{grad}) simply selects the
effective periods with which the coupling oscillates.

Note that for $\omega = E_F$, each wave vector
$q_j^s({\bf q}_\parallel,E_F)$ represents a surface in the reciprocal
space and when put together, the surfaces $q_j^s$ reproduce the spacer
Fermi surface. Following this argument and bearing in mind that
non-real values of $q_j^s$ strongly damp the oscillations, we stress
that only real solutions must be taken into account. 
The possible stationary solutions of Eq.(\ref{grad}) reflect the lattice
structure and are usually located at points of high symmetry, even though
they may occur at general points of the 2-dimensional BZ.

In such limit we obtain an analytical expression for the coupling given
by\cite{ed1,mt1,zeetal}
\begin{equation}
\Delta \Omega(\pi) = \sum_\alpha\,\sum_{(m_1^1,...,m_j^s,...)}
{\cal K}_{(m_1^1,...,m_j^s,...)}^\alpha \,\,e^{i \, n \, d \, {\bar q}}\,,
\label{jspa} 
\end{equation}
\begin{equation}
{\cal K}_{(m_1^1,...,m_j^s,...)}^\alpha = \left({ 2 \sqrt{2} d \over \pi
\beta n}\right) {\sigma_{(m_1^1,...,m_j^s,...)}^\alpha
C_{(m_1^1,...,m_j^s,...)}^\alpha \vert 
{\partial^2 \bar q \over \partial q_x^2} {\partial^2 \bar q \over
\partial q_y^2} \vert^{-1/2} \over \sinh\left[\left({\partial \bar q
\over \partial \omega}\right)_{E_F} {\pi n d \over \beta}\right]} \,\,,
\label{Tspa}
\end{equation}
where $\bar q = \sum_{s j} (m_j^s \, {q_j^s}^\alpha)$,
${q_j^s}^{\alpha} = q_j^s({\bf q}_\parallel^\alpha,E_F)$, and $\beta =
1/k_B T$. Here we neglect the energy dependence of the Fourier
coefficients which in some cases \cite{spa2} can be important. The
factor $\sigma_{(m_1^1,...,m_j^s,...)}^\alpha$ takes the 
value $i$ for a minimum, $-i$ for a maximum and $1$ for a saddle point
of the surface ${\bar q}(q_x,q_y)$. The sum over $\alpha$ regards all
different stationary solutions and the inner sum over
${(m_1^1,...,m_j^s,...)}$ runs over the values allowed by the stationary
phase condition. The coupling oscillates with periods
$\lambda_{(m_1^1,...,m_j^s,...)}^{\alpha} = 2 \pi / \vert \sum_{s
j}(m_j^s {q_j^s}^{\alpha}) \vert$ and their
weights depend on geometric factors as well as on the Fourier
coefficients. 

An important issue is the relationship between the Fermi
surface and the wave vectors that effectively contribute to the
coupling. The lowest order combinations amongst the possible wave
vectors $q_j^s({\bf q}_\parallel,E_F)$ bear a direct correspondence with
the FS. Indeed, this correspondence agrees with the RKKY geometrical
picture for selecting the oscillation periods of the interlayer
coupling. However, in the general case of a higher order combination of
wave vectors the RRKY relation no longer holds, and the
effective wave vectors are not directly related to the spacer FS. This
is a very important result which shows that the geometrical criteria used
for selecting the oscillation periods of the interlayer coupling are
incomplete. Even though the wave vectors $q_j^s$ are directly related to
the FS, the general wave vectors $\sum_{s j} (m_j^s \, q_j^s)$ are not. 
Furthermore, note that if $g_{n 0}({\bf q}_\parallel^\alpha,E_F)$ has
only one oscillation period, the full expansion in Eq. (\ref{jspa})
contains merely harmonics of the fundamental RKKY periods. In other
words, if the FS is cut in a single point by the line drawn from the
origin perpendicular to the layers for a fixed ${\bf q}_\parallel$, the
RKKY periods and its harmonics are the only ones present in the
coupling. In fact, it explains the agreement between the periods
predicted by the RKKY and QWT in the models previously
mentioned\cite{ed1,ed2,mt1,fcc111}. For a multi-periodic $g_{n 0}({\bf
q}_\parallel,E_F)$ though, where the FS is intersected in more than one
point corresponding to incommensurate wave vectors, the higher order
combinations of the wave vectors correspond neither to a RKKY period nor
to any of its harmonics, but to {\it fundamentally new periods}. 

Multi-periodicity of the off-diagonal propagator is a necessary but not
sufficient condition. The argument above simply shows that non-RKKY
behaviour is possible in presence of multi-periodic $g_{n
0}$. To find out whether it really occurs one has to solve
Eq.(\ref{grad}) and check whether there are real wave vectors $q_j^s$
for values of $m_1^1$,...,$m_j^s,...$ other than the ones obtained by
the RKKY theory.  

In summary, the determination of the oscillation periods results from
the analysis of the bulk spacer off-diagonal propagators where the
relevant real wave vectors, under the stationary phase condition, are
obtained from the band structure. The relation between the spacer FS and
the oscillation periods is the same as the geometrical RKKY picture for
the lowest order terms. Nevertheless, for higher order contributions a
direct correspondence with the FS is not evident and the stationary
phase condition displayed in Eq.(\ref{grad}) determines the periods
with which the coupling oscillates. We have shown that oscillation periods
other than the ones predicted by the RKKY theory may arise.
To illustrate the method and unambiguously prove the existence and
relevance of periods other than RKKY ones, we next evaluate the particular
case of the FCC lattice along the (110) direction within the single-band 
tight-binding model. 

\section{Applying the method}
\label{fcc110}

Having presented the general method for determining the periods of the
interlayer coupling, here we apply it to the particular case of a
FCC(110) system within the single-band nearest-neighbour tight-binding
model, which unmistakably shows the existence of non-RKKY periods.
Moreover, in the light of the QWT this system is interesting because of
the difficulty in determining analytically the energies of the
resonances and size quantized states in such a structure\cite{harrison}.
For a single band the notation introduced in the previous section
becomes simpler due to the omission of the band index $s$. In addition,
all the matrices over orbital indices become scalars.

\begin{figure}
\begin{center}\mbox{\psboxscaled{800}{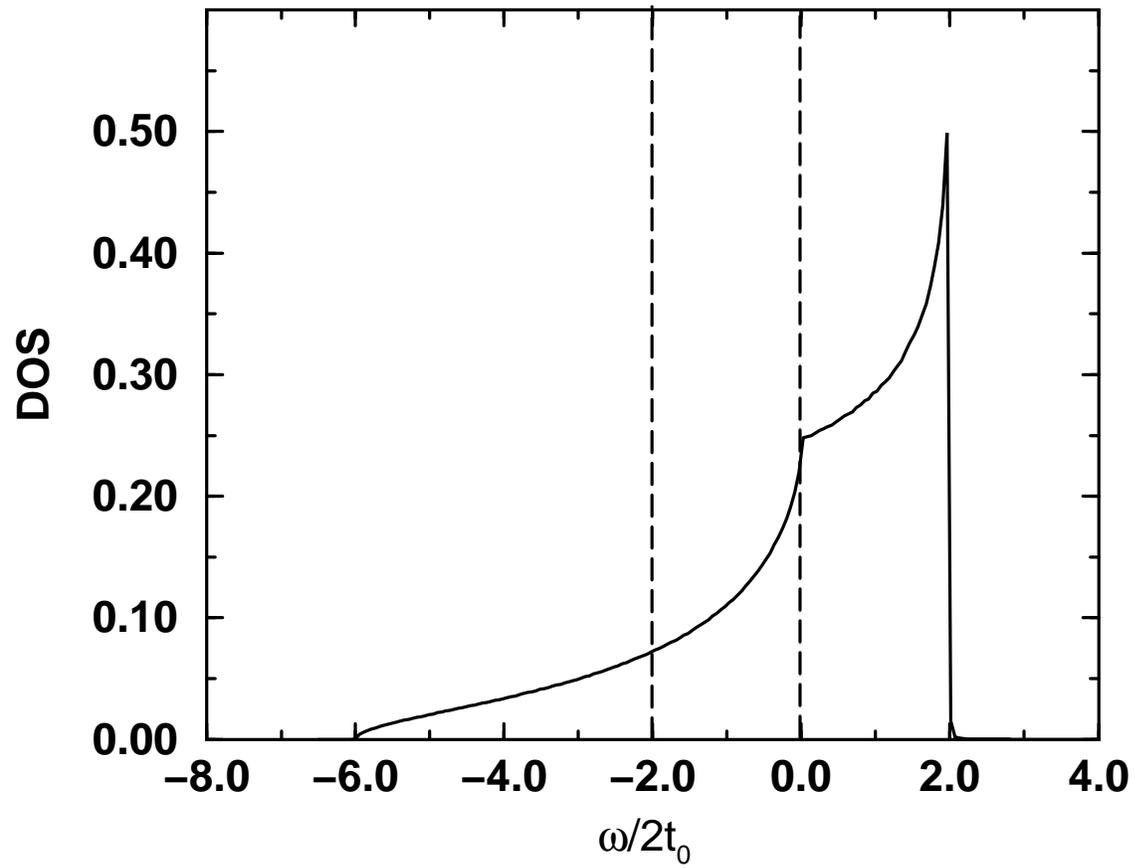}}\end{center}
\caption{Density of states of the bulk FCC tight-binding model. The
vertical dashed lines separate the three distinct regions.}
\label{dos}
\end{figure}

We label the nearest neighbour hopping $t_0$ and with no loss of
generality assume the on-site potential $\epsilon_0 = 0$. The bulk
density of states of the FCC lattice is displayed in Fig.\ref{dos}. 
It is worth pointing out that there are different contributions to the
interlayer coupling, depending on the position of the Fermi energy
inside the band. Three distinct regions are observed and their
boundaries are shown by the vertical dashed lines in Fig.\ref{dos}. In
addition to the traditional polyhedral BZ, we make use of an auxiliary
prismatic BZ due to the 2-dimensional symmetry of the
multilayers\cite{bruno2}. Both zones are equivalent and enclose the same
volume in reciprocal space. The energy boundaries indicated correspond
to Fermi energies such that the FS just touches the two distinct
Brillouin zones. More specifically, for $E_F = -4 t_0$ the corresponding
FS touches the prismatic BZ, whereas $E_F = 0$ is the exact Fermi energy
where the FS opens the necks which are familiar from the noble metals
\cite{ashcroft}.

As previously mentioned, even within the nearest neighbour tight-binding
model, in this case electrons are capable of hopping between nearest
and next nearest planes. The bulk band structure is then given by
\begin{equation}
E({\bf q}_\parallel,q_\perp) = \epsilon({\bf q}_\parallel) + 2 \,t_1({\bf
q}_\parallel) \cos(q_\perp \,d) + 2 \,t_2({\bf q}_\parallel) \cos(2
q_\perp \,d) \,\,. 
\label{E}
\end{equation}
We assume the $z$ direction perpendicular to the layers. The
2-dimensional BZ over which the stationary solutions must be searched
for is a rectangle defined by $-\pi/2d \le q_x \le \pi/2d$ and $-\pi
\sqrt{2}/4d \le q_y \le \pi \sqrt{2}/4d$, where $q_x$ and $q_y$ are the
components of ${\bf q}_\parallel$. The explicit dependence of $E$ on
${\bf q}_\parallel$ is given by 
$\epsilon({\bf q}_\parallel) = -2 \,t_0 \cos(2 q_x d)$, $t_1({\bf
q}_\parallel) = -4 \,t_0 \cos(q_x d) \cos(\sqrt{2} q_y d)$ and 
$t_2({\bf q}_\parallel) = - t_0$.

\begin{figure}
\begin{center}\mbox{\psboxscaled{700}{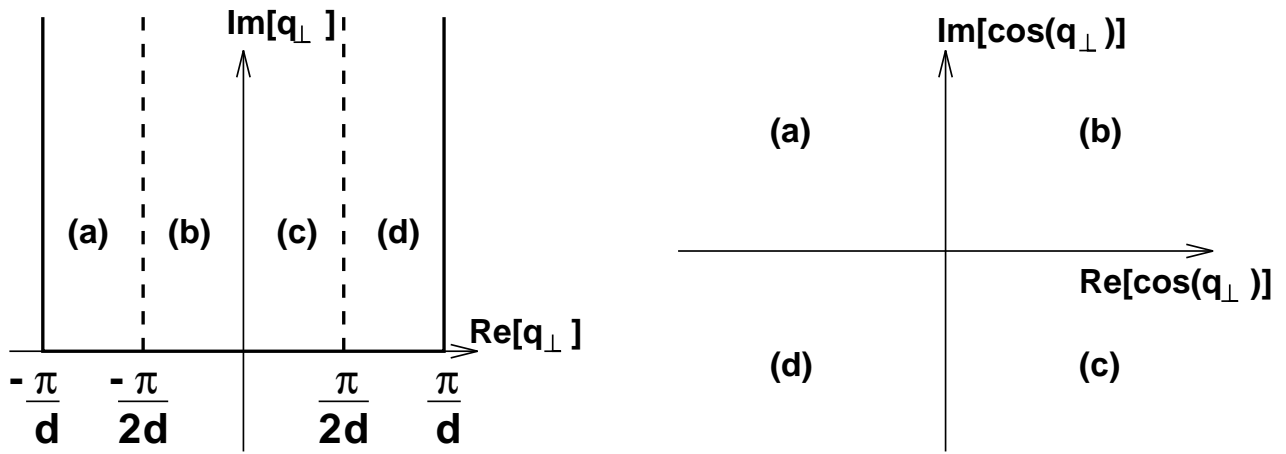}}\end{center}
\caption{Correspondence between the $q_\perp$ and
$\cos(q_\perp)$-complex planes for $\ell < m$. By considering $q_\perp$
inside the integration contour (bold path on the left), each point on
one plane has an unique correspondence on the other one. For the case
$\ell > m$, since $\cos(q_\perp)$ is even, the map of the
$q_\perp$-complex plane must be reflected about the axes.} 
\label{map}
\end{figure}

Following the steps of the previous section, $\omega^+ - E({\bf
q}_\parallel,q_\perp) = 0$ gives the poles which yield the periods of
$g_{n 0}$. Eq.(\ref{E}) can be rewritten in terms of a second degree
polynomium in $\cos(q_\perp d)$ whose solutions $q_1$ and $q_2$ are
given by 
\begin{equation}
\cos(q_j \,d) = - \left\{ {\gamma + (-1)^j \sqrt{\gamma^2 +
{8\left(\omega^+ - \epsilon + 2 t_2\right) \over 2 t_2}} \over 4}
\right\}\;,
\label{cosq}
\end{equation}
where $\gamma({\bf q}_\parallel) = t_1/t_2$ and $j$ is either 1 or 2.
Nevertheless, the poles $q_1$ and $q_2$ are in the $q_\perp$-complex
plane. By mapping the relationship between the two 
complex functions we can uniquely relate one value of $q_\perp$ inside
the integration contour to the corresponding value of $\cos(q_\perp d)$
and vice versa, as shown in Fig.\ref{map}. The equation above shows that
$\cos(q_1 \,d)$ and $\cos(q_2 \,d)$ have opposite imaginary parts and
therefore, following the maps in Fig.\ref{map}, $q_1$ and $q_2$ have
opposite real parts. This confirms what has been said in the previous
section concerning the position of two different wave vectors for a
given band $s$. 

In addition, looking at Eq.(\ref{cosq}) with $\omega = E_F$ it is clear
that non-RKKY behaviour cannot arise in the bottom region of the band 
($-12 t_0 < E_F < -4 t_0$). This is due to the fact that in such a
region $q_1$ and $q_2$ cannot both be real.
In fact, the corresponding FS in this region is spherical-like. However,
in the range $-4 t_0 < E_F < 4 t_0$, corresponding to the top regions
inside the band, both $q_1$ and $q_2$ may be real indicating that
non-RKKY periods are possible. 

Having obtained the poles, from Eqs.(\ref{glm}) and (\ref{AA}) we have 
\begin{equation}
g_{n 0}({\bf q}_\parallel,\omega) = A_1({\bf q}_\parallel,\omega)
\,\,e^{i q_1({\bf q}_\parallel,\omega) n d} + A_2({\bf
q}_\parallel,\omega) \,\,e^{i q_2({\bf q}_\parallel,\omega) n d} \,\,,
\label{g0n2}
\end{equation}
where 
\begin{equation}
A_j({\bf q}_\parallel,\omega) = \left\{2i\left[\cos(q_1 d) -
\cos(q_2 d)\right]\sqrt{1 - \cos^2(q_j d)}\right\}^{-1} \,\,.
\end{equation}
In the present case $g_{0 n} = g_{n 0}$. Eq.(\ref{g0n2}) shows
that for values of ${\bf q}_\parallel$ and $\omega$, for which $q_1$ and
$q_2$ are real, $g_{n 0}$ oscillates with the superposition of two
periods, $2 \pi d/\vert q_1\vert$ and $2 \pi d/\vert q_2\vert$, which
are in general incommensurate. In those cases $g_{n 0}$ exhibits a
quasi-periodic dependence on $n$.

Because there are only two poles in this case, the function ${\cal F}$, 
following Eq.(\ref{f}), becomes
\begin{equation}
{\cal F} = \sum_{m_1,m_2}\, C_{m_1,m_2} \,\, e^{i(m_1 q_1 + m_2 q_2) n
d} \,\,,
\label{f2}
\end{equation}
where the Fourier coefficients are given by Eq.(\ref{coefs}).
Within the single-band model, the Fourier coefficients can be calculated
in an alternative way, shown in Appendix \ref{fourier}. Following the
Appendix, we recall that the lowest order Fourier coefficients
correspond to the RKKY-limit ${\cal F}_{RKKY}$. They are $C_{m_1,m_2}$
where $m_1 + m_2 = 2$, $\vert m_1 \vert \le 2$ and $\vert m_2 \vert \le
2$. More specifically, $C_{2,0} = {{\cal T} A_1^2 \over 2 i}$, 
$C_{0,2} = {{\cal T} A_2^2 \over 2 i}$, $C_{1,1} = {{\cal T} A_1 A_2
\over i}$, and the property $C_{-m_1,-m_2} = (C_{m_1,m_2})^*$ define all
the six coefficients present in the RKKY expression, where ${\cal T} = 4
\tau_{\uparrow}\tau_{\downarrow}\,(1-\tau_{\uparrow}\,g_{0
0})(1-\tau_{\downarrow}\,g_{0 0})$ and $\tau_{\uparrow \downarrow}$ are
defined after Eq.(\ref{G}). Beyond the lowest order
coefficients are the terms expected to contribute to the non-RKKY
behaviour. It is then illustrative to present one of these terms, namely 
$C_{3,1}$,
\begin{equation}
C_{3,1} = {2 {\cal T} A_1^3 A_2 \over i} \left[
\left(\tau_\uparrow\right)^2 + \left(\tau_\downarrow\right)^2 - {{\cal
T} \over 2}\right] \,\,. 
\end{equation}

The stationary phase condition in Eq.(\ref{grad}) is rewritten as
\begin{equation}
m_1 {{\bf \nabla}_\parallel} \,q_1({\bf q}_\parallel,E_F) + m_2 {{\bf 
\nabla}_\parallel} \,q_2({\bf q}_\parallel,E_F) = 0 \,\,.
\label{grad2}
\end{equation}
The condition above imposes that both gradients must be colinear and
moreover, the ratio between their moduli must be a rational number. The
determination of the stationary points ${\bf q}_\parallel^{\it \alpha} =
(q_x,q_y)$ yielding the periods $\lambda_{m_1,m_2}^{\alpha} = 2 \pi /
\vert m_1 \, q_1^\alpha + m_2 \, q_2^\alpha \vert$ is rather tedious but
straightforward. As previously mentioned, the solutions are often
located at highly symmetric points of the BZ but more general solutions
may occur. Nevertheless, because the RKKY periods as well as the
most important higher order terms come only from those high-symmetry
points, we shall firstly focus on these solutions. 
Across the whole energy band, four stationary points are found on the
corners of the 2-dimensional irreducible BZ. They are ${\bf
q}_\parallel^{\it a} = (0,0)$, ${\bf q}_\parallel^{\it b} = (0,\pm \pi
\sqrt{2}/4d)$, ${\bf q}_\parallel^{\it c} = (\pm \pi/2d,\pm \pi
\sqrt{2}/4d)$ and ${\bf q}_\parallel^{\it d} = (\pm \pi/2d,0)$. 

\begin{figure}
\begin{center}\mbox{\psboxscaled{800}{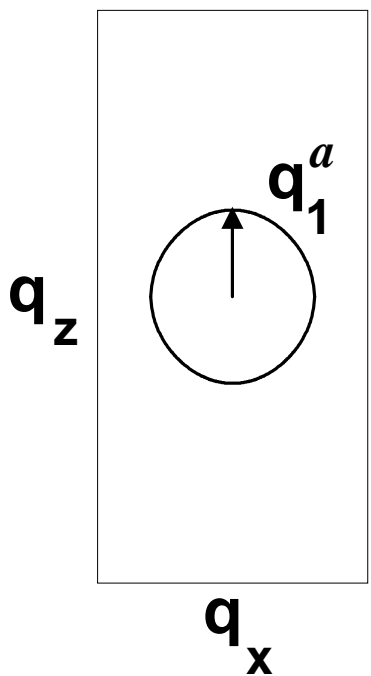}}\end{center}
\caption{Cross section of the Fermi surface for $E_F = -3.0$
(Region 1) and $q_y = 0$. The rectangle is a cross section of the
prismatic Brillouin zone and the wave vector $q_1^{\it a}$ is also
displayed. } 
\label{fsef_3}
\end{figure}

In the first region $\lambda_{2,0}^{\it a}$ is the only period,
which is actually present along the entire band. The graphical
relation  between the spacer FS and the oscillation period is displayed
in Fig.\ref{fsef_3}, where the FS for an arbitrary energy ($E_F/2 t_0 = 
-3.0$) is shown together with the wave vector $q_1^{\it a}$ and the
cross section of the prismatic BZ for $q_y = 0$. The cross section is
defined by $-\pi/2d \le q_x \le \pi/2d$ and $-\pi/d \le q_z \le \pi/d$. 
Note that $2 {\bf q}_1^{\it a}$ corresponds to the extremal wave vector
spanning the FS, in according to the RKKY picture. Clearly, higher order
periods $\lambda_{m_1,0}$ for $m_1 > 2$ are merely harmonics of the
fundamental $\lambda_{2,0}$.  

\begin{figure}
\begin{center}\mbox{\psboxscaled{800}{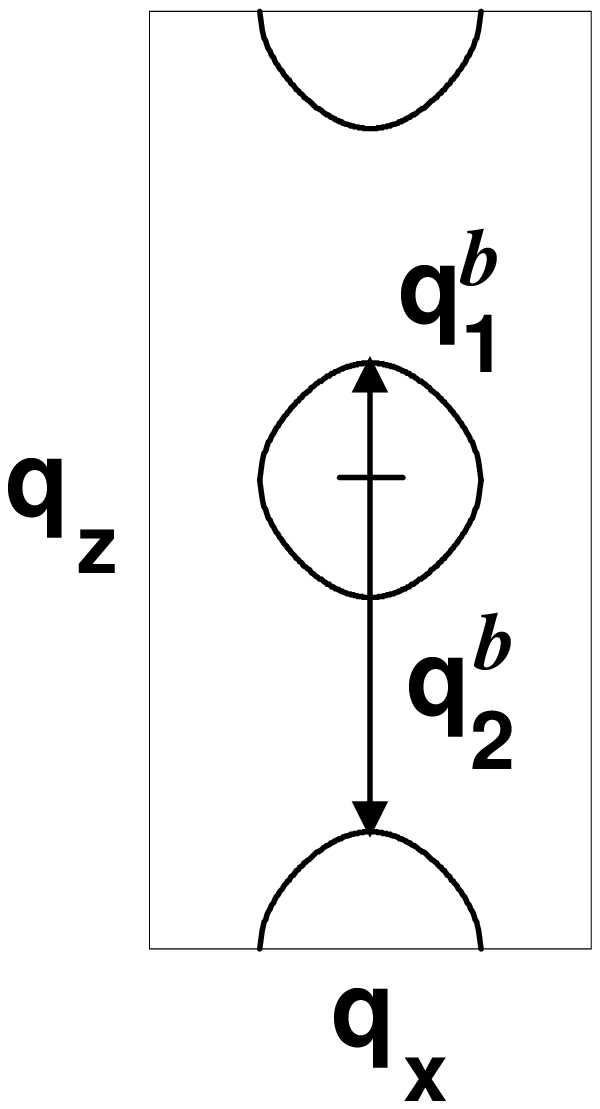}}\end{center}
\caption{Cross section of the Fermi surface for $E_F = -1.0$
(Region 2) and $q_y = \pi\protect\sqrt{2}/4d$. The rectangle is a
cross section of the prismatic Brillouin zone. The wave vectors
$q_1^{\it b}$ and $q_2^{\it b}$ are also displayed.}
\label{fsef_1}
\end{figure}

In the second region an additional period $\lambda_{1,1}^{\it b}$
arises. Similarly, the vector $q_1^{\it b} + q_2^{\it b}$ yields the
fundamental period and possible higher combinations 
$m_1(q_1^{\it b} + q_2^{\it b})$ add no new terms but harmonics.
Fig.\ref{fsef_1} shows a cross section of the FS ($E_F/2 t_0 = 
-1.0$) for $q_y = \pi \sqrt{2}/4d$. The wave vectors ${\bf q}_1^{\it b}$
and ${\bf q}_2^{\it b}$ are highlighted and it is clear that ${\bf
q}_1^{\it b} + {\bf q}_2^{\it b}$ satisfies the RKKY criteria. At first
glance one could think, accordingly to Fig.\ref{fsef_1}, that $2
{\bf q}_1^{\it b}$ and $2 {\bf q}_2^{\it b}$ are also contributory wave
vectors for the same reasons as in the first region. Nevertheless, by
looking at the $q_y$-dependence of the FS we find that those vectors are
not solutions of Eq.(\ref{grad2}).

\begin{figure}
\begin{center}\mbox{\psboxscaled{800}{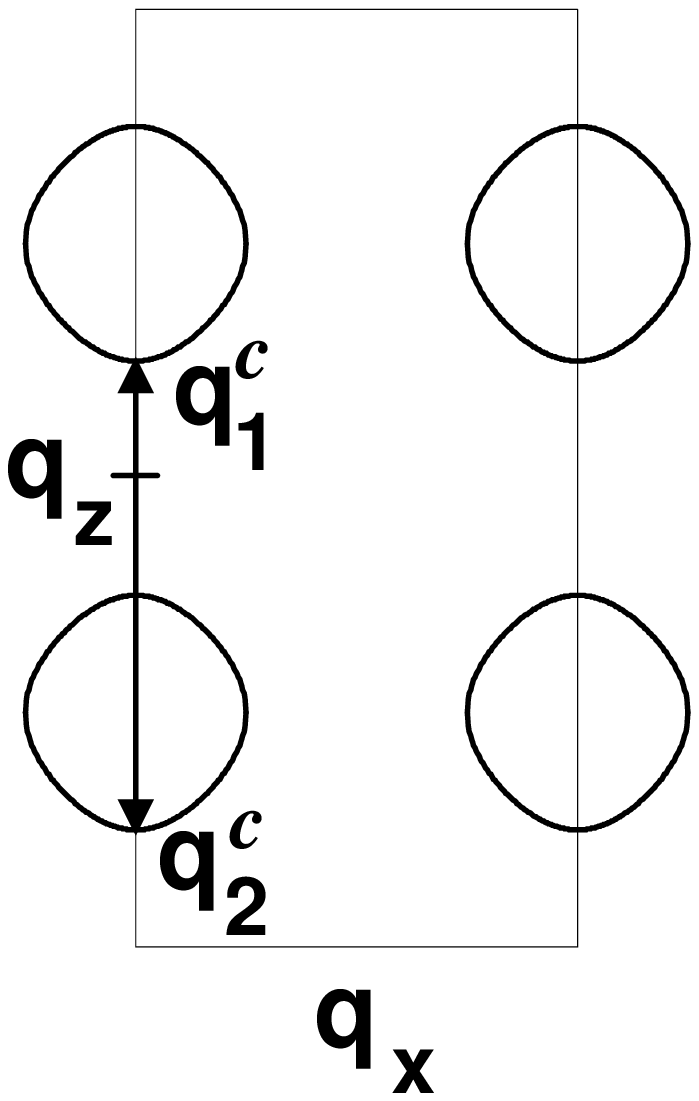}}\end{center}
\protect\caption{Cross section of the Fermi surface for $E_F = 1.0$
(Region 3) and $q_y = \pi\protect\sqrt{2}/4d$. The rectangle is a cross
section of the prismatic Brillouin zone. The wave vectors $q_1^{\it c}$
and $q_2^{\it c}$ are also displayed.}
\label{fsef1}
\end{figure}

The most interesting case occurs in the top region, where the FS has
necks resembling those of the noble metals. In addition to
$\lambda_{2,0}^{\it a}$ at ${\bf q}_\parallel^{\it a}$, other stationary
solutions occur at ${\bf q}_\parallel^{\it c}$ and ${\bf
q}_\parallel^{\it d}$. At ${\bf q}_\parallel^{\it d}$, analogously to
the second region, the solution comes from ${\bf q}_1^{\it d} + {\bf
q}_2^{\it d}$ and does not introduce new periods. At ${\bf
q}_\parallel^{\it c}$ though, both gradients in
Eq.(\ref{grad2}) vanish simultaneously allowing any values of $m_1$ and
$m_2$ to satisfy the stationary phase condition. Therefore, the coupling
oscillates with as many periods $\lambda_{m_1,m_2}^{\it c}$ as
there are non-zero coefficients $C_{m_1,m_2}^{\it c}$.
The lowest order $\lambda_{2,0}^{\it c}$, $\lambda_{0,2}^{\it c}$
and $\lambda_{1,1}^{\it c}$ correspond to the RKKY periods and the
geometrical relation between the wave vectors ${\bf q}_1^{\it c}$, ${\bf
q}_2^{\it c}$ and the FS in Fig.\ref{fsef1} confirms this point.
For the present model we find that $q_1^{\it c} -  q_2^{\it c} = \pi$.
Thus, the oscillation periods $\lambda_{2,0}^{\it c}$ and
$\lambda_{0,2}^{\it c}$ cannot be distinguished just by looking at
discrete integer values of $n$. In addition, $\lambda_{1,1}^{\it c} =
\lambda_{1,1}^{\it d}$, which leads the third energy region to contain
three non-equivalent RKKY periods. We recall that in FCC Cu $E_F$ lies
in the third energy region with FS necks. A quantitative description of
the oscillation periods for Cu can be obtained within the present
framework by going beyond nearest neighbours and using the tight-binding 
parameters of Halse\cite{halse}. Then $\lambda_{1,1}^{\it c}$ and
$\lambda_{1,1}^{\it d}$ become distinct periods and, by taking into
account the equivalence of $\lambda_{2,0}^{\it c}$ and
$\lambda_{0,2}^{\it c}$ for a discrete lattice, we find exactly the four
RKKY periods of Bruno and Chappert \cite{bruno1}.

However, because in general the wave vectors ${\bf q}_1^{\it c}$ and
${\bf q}_2^{\it c}$ are incommensurate, higher order periods are not
simple harmonics of the RKKY ones, unlike in the two lower regions. In 
fact, the period $\lambda_{3,1}^{\it c}$ does not correspond to any
harmonics of the fundamental RKKY periods. Furthermore, even though the
wave vectors ${\bf q}_1^{\it c}$ and ${\bf q}_2^{\it c}$ separately have 
clear correspondences with the FS, as shown in Fig.\ref{fsef1}, the wave
vector $3 {\bf q}_1^{\it c} + {\bf q}_2^{\it c}$ has not. From this
argument it becomes clear that higher order periods do not satisfy the
geometrical picture of the RKKY theory, having instead an indirect relation
between the wave vectors and the FS. 

\begin{figure}
\begin{center}\mbox{\psboxscaled{800}{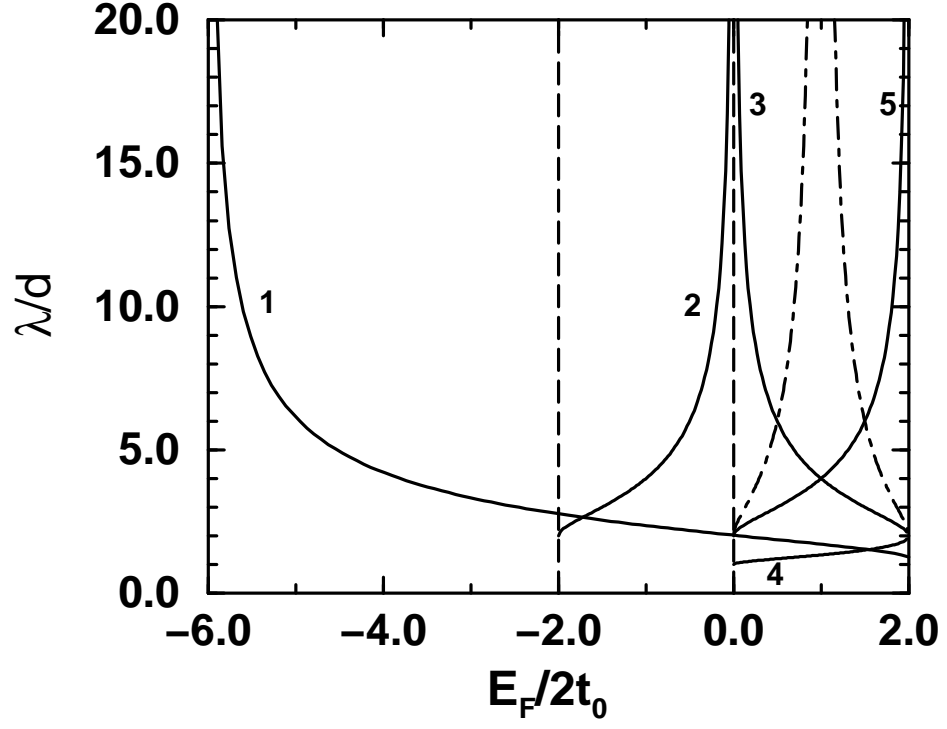}}\end{center}
\protect\caption{Oscillation periods of the interlayer coupling as
a function of the Fermi energy $E_F$. The vertical dashed lines separate
the three distinct regions inside the band. The full lines
labeled from $1$ to $5$ are the RKKY periods (see text) whereas the
dot-dashed line in the top region is one of the non-RKKY periods, namely
$\lambda_{3,1}^{\it c}$. }
\label{periods}
\end{figure}

\begin{figure}
\begin{center}\mbox{\psboxscaled{800}{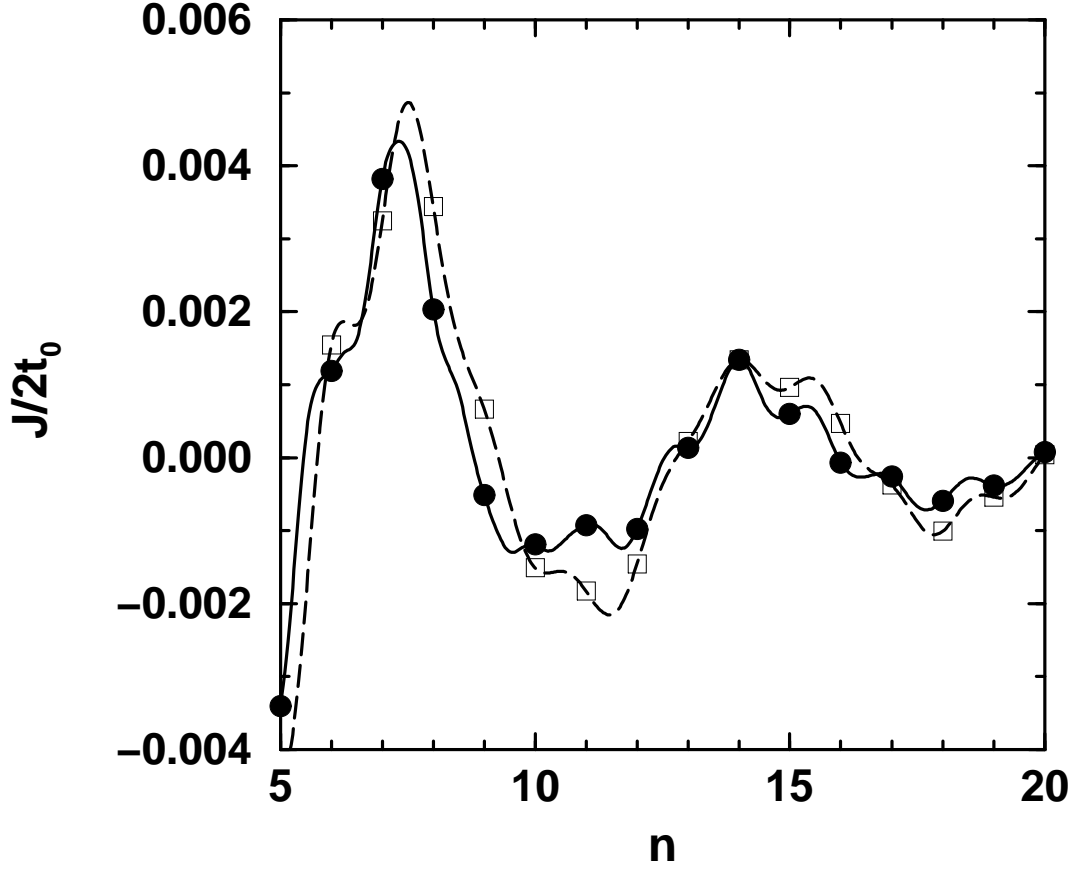}}\end{center}
\protect\caption{ Full line shows the interlayer coupling for $E_F/2
t_0 = 1.64$, $V_{ex} = 0.15 W$ and $k_BT = 2.0 \times 10^{-3} W$,
whereas the dashed line is the RKKY prediction for the same parameters
scaled down by a factor of 8 for comparison. }
\label{result}
\end{figure}

Fig.\ref{periods} summarizes the results obtained here by displaying the
oscillation periods as a function of the Fermi energy $E_F$. The full
lines labeled from 1 to 5 are the fundamental RKKY periods 
$\lambda_{2,0}^{\it a}$, $\lambda_{1,1}^{\it b}$, $\lambda_{2,0}^{\it
c}$, $\lambda_{0,2}^{\it c}$ and $\lambda_{1,1}^{\it c}$, respectively,
and the dot-dashed line in the top region represents only one of the
non-RKKY periods, namely $\lambda_{3,1}^{\it c}$. Recalling the relation
$q_1^{\it c} - q_2^{\it c} = \pi$, we see that the oscillation periods 
$\lambda_{2,0}^{\it c}$ and $\lambda_{0,2}^{\it c}$ cannot be
distinguished just by looking at discrete integer values of $n$. Still
focusing on the top energy region, Fig.\ref{result} exhibits the
bilinear coupling $J_1 = \Delta \Omega(\pi)/2$ at temperature $T = 2.0
\times 10^{-3} W/k_B$ as a function of the spacer thickness for
$E_F/2t_0 = 1.64$ and $V_{ex} = 0.15W$, where $W = 16t_0$ is the spacer
band-width. This value of $E_F$ is chosen so that $\lambda_{3,1}^{\it
c}$ is well separated from the RKKY periods, as may be seen in
Fig.\ref{periods}. The full line in Fig.\ref{result} corresponds to the
result obtained from Eqs.(\ref{j}) and (\ref{Fwq}), whereas the dashed
line is the RKKY approximation, following Eqs.(\ref{j}) and
(\ref{Frkky}), scaled down by a factor of 8 for reasons of comparison.
Despite the agreement in the dominant long period, the difference in the
fine structure of the oscillations reflects contributions beyond the
fundamental RKKY periods.

\begin{figure}
\begin{center}\mbox{\psboxscaled{800}{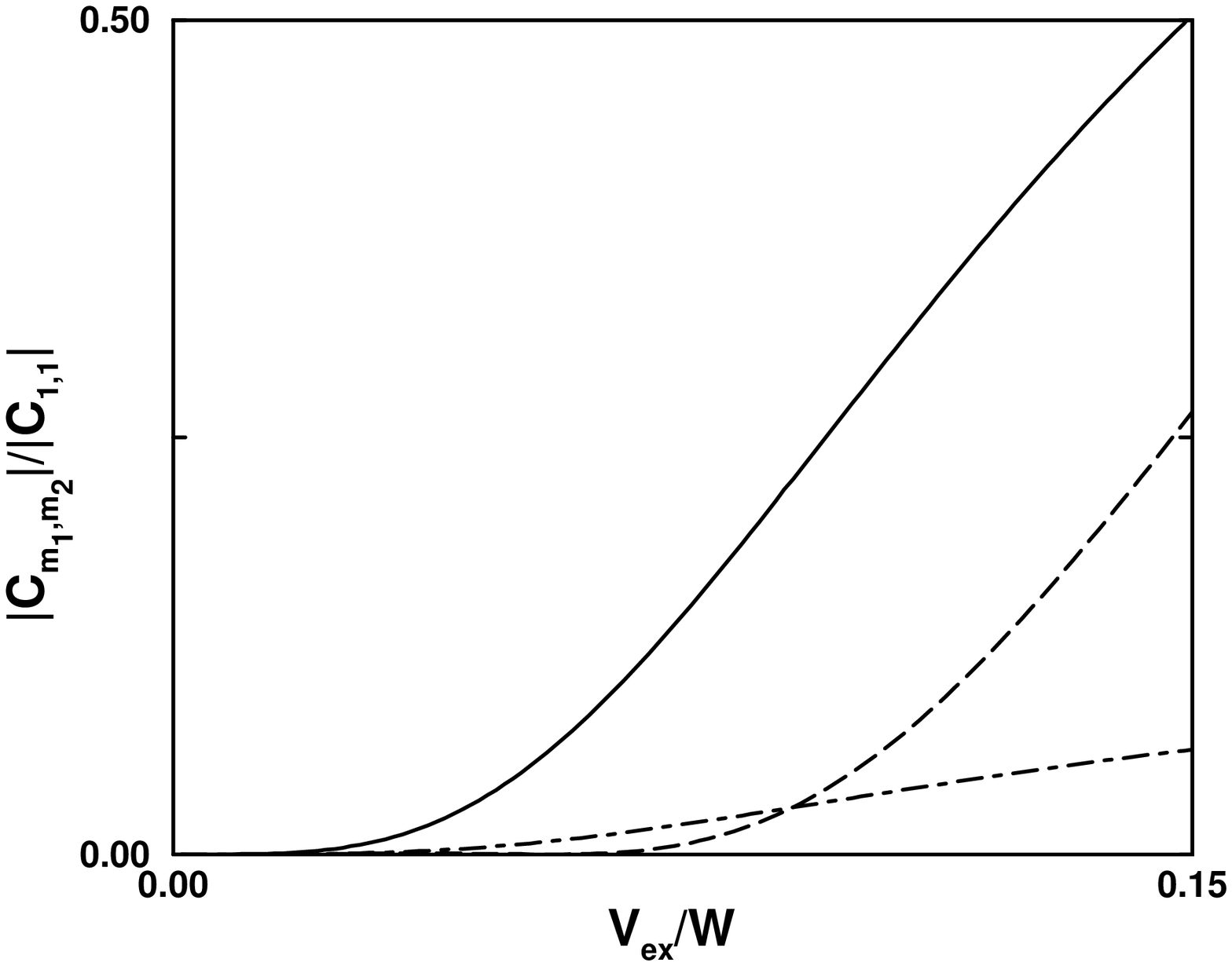}}\end{center}
\protect\caption{Ratios $\vert C_{3,1}^{\it c} \vert / \vert
C_{1,1}^{\it c} \vert$ (full line), $\vert C_{4,2}^{\it c} \vert / \vert 
C_{1,1}^{\it c} \vert$ (dashed line) and $\vert C_{4,0}^{\it c} \vert /
\vert C_{1,1}^{\it c} \vert$ (dot-dashed line) as a function of
$V_{ex}$. }
\label{coefsfig}
\end{figure}

\begin{figure}
\begin{center}\mbox{\psboxscaled{800}{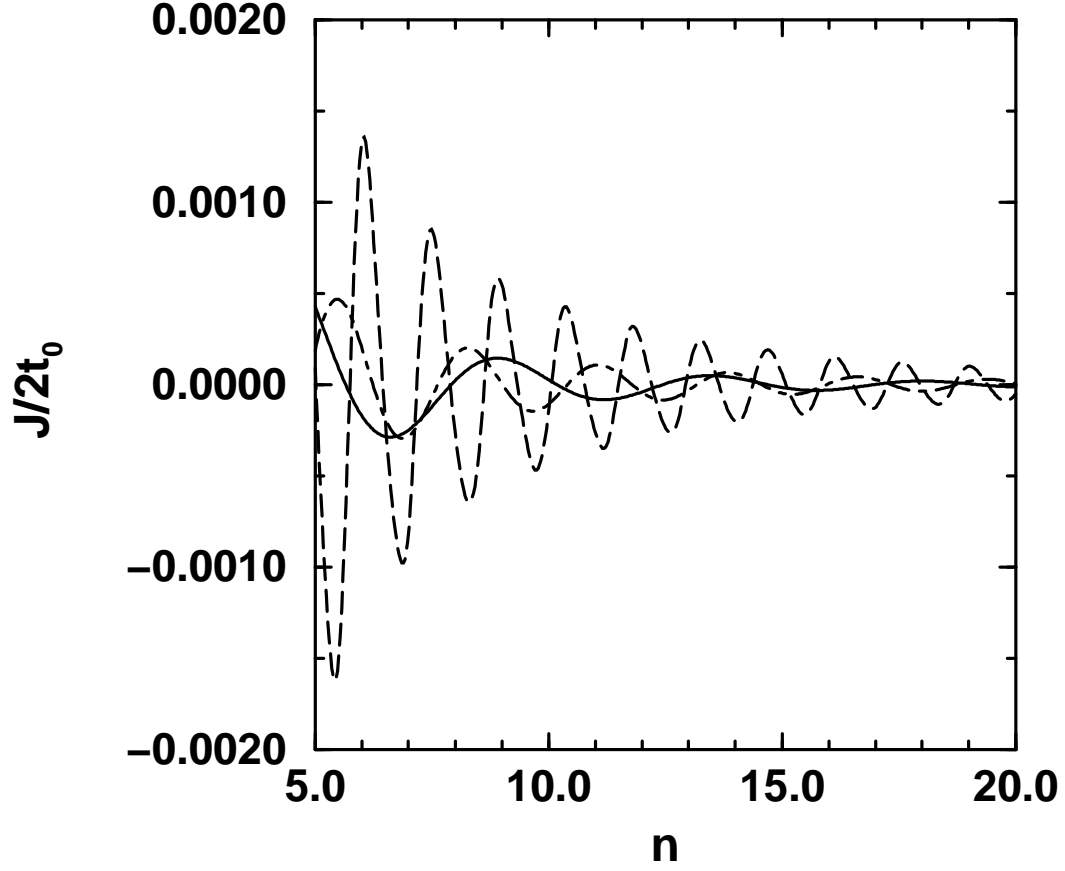}}\end{center}
\protect\caption{Separate contributions to the interlayer coupling
through the stationary phase approximation. The full line is
$\lambda_{3,1}^{\it c}$, dashed line is $\lambda_{2,0}^{\it a}$ and
dot-dashed line is $\lambda_{2,0}^{\it c}$. $E_F/2 t_0 = 1.64$, $V_{ex} =
0.15 W$ and $k_b T = 2.0 \times 10^{-3} W$.}
\label{spa}
\end{figure}

The evaluation of the stationary phase approximation in this case
unmistakably proves the existence and relevance of the non-RKKY periods.
Fig.\ref{coefsfig} displays the absolute value of the ratio between some
Fourier coefficients $C_{m_1,m_2}^{\it c}$ and $C_{1,1}^{\it c}$, the
largest coefficient of a fundamental RKKY period, as a function of
$V_{ex}$. The full line corresponds to the Fourier coefficient
associated with the new period $\lambda_{3,1}^{\it c}$ whereas the
dashed and dot-dashed lines correspond to $\lambda_{4,2}^{\it c}$ and
$\lambda_{4,0}^{\it c}$, respectively, which are simple harmonics of the 
fundamental RKKY periods. As expected for very small exchange
splittings, the magnitudes of higher order coefficients relative to that
of the fundamental RKKY ones are negligible\cite{paper1}, indicating that
non-RKKY behaviour becomes more important as the exchange splitting is
increased. An interesting consequence is that, even though the periods
are determined from the spacer FS, their relevances depend on the
nature of the magnetic materials involved. 
Note that for $V_{ex} \approx 0.15 W$ the ratio $\vert
C_{3,1}^{\it c} \vert / \vert C_{1,1}^{\it c} \vert \approx 1/2$. This 
indicates that for such a value of $V_{ex}$, which is exactly the same as
in Fig. \ref{result}, the Fourier coefficients are of the same order of
magnitude; the only reason why both periods do not have comparable
weights is due to geometrical factors determined by the curvatures of
the respective surfaces. In fact, the curvature of the surface $q_1 +
q_2$ vanishes at ${\bf q}_\parallel^{\it c}$, making it impossible to
evaluate the stationary phase approximation for the period
$\lambda_{1,1}^{\it c}$ even though it undoubtedly indicates a strong
contribution. Taking into account the geometrical curvatures when
possible, we evaluate the stationary phase approximation and show in
Fig.\ref{spa} the separate contribution of some periods to the exchange
coupling. The full line is the new period $\lambda_{3,1}^{\it c}$
whereas the dashed line and the the dot-dashed line represent
$\lambda_{2,0}^{\it a}$ and $\lambda_{2,0}^{\it c}$, respectively.
Clearly, $\lambda_{3,1}^{\it c}$ is as important as the RKKY periods,
except for the long dominant one. It is worth pointing out that the rate
of decay of the oscillation amplitude associated 
with the new periods is exactly the same as for the lowest order ones,
falling off as $1/n^2$ for $T = 0$. Concerning the actual amplitude of
the oscillations, two points should be highlighted. The strength of
all periods depends rather critically on the matching of the bands in
real materials, which is not well reproduced by our single-orbital
model, and the omission of the energy dependence of the Fourier
coefficients could be serious for Cu-based systems \cite{spa2}. 

The solutions treated above all come from the high-symmetry points ${\bf 
q}_\parallel^{\it a}$, ${\bf q}_\parallel^{\it b}$, ${\bf
q}_\parallel^{\it c}$ and ${\bf q}_\parallel^{\it d}$. Note that
fundamentally new periods arise but always in the presence of RKKY ones.
They result from crossed mixtures of the fundamental wave vectors,
which in turn yield the RKKY periods. Nevertheless in general, as previously
stated, solutions of Eq.(\ref{grad2}) away from high-symmetry points are
found and we point out that their contributions are not necessarily
associated with the existence of RKKY periods. In fact, it can be shown
in the second energy region that energy-dependent solutions
exist along the line $q_x = 0$ with corresponding wave vectors ${\bf q}_1$
and ${\bf q}_2$ which do not yield any possible RKKY period. 
This indicates there are colinear gradients along that line satisfying
Eq.(\ref{grad2}) for different values of $m_1$ and $m_2$ even though no
RKKY stationary solutions are obtained. These solutions are found to be
far less significant in their contribution to the exchange coupling than
the highly symmetric ones.

\section{Conclusions}
\label{conclusions}

In summary, we have shown that the RKKY theory is not capable of
predicting all the oscillation periods of the interlayer coupling, as
generally believed. The actual relation between the periods and the
spacer FS is more subtle than the RKKY geometrical picture. Even though
the lowest order periods indeed correspond to the RKKY predictions,
non-negligible higher order terms arise and do not bear any direct
correspondence to the FS. Instead, they correspond to indirect
combinations of fundamental wave vectors, which in turn, may or may not
yield RKKY periods.  The appearance of non-RKKY oscillations clearly
requires the multi-periodicity of the bulk spacer off-diagonal
propagator as a function of the spacer thickness, although this is not a
sufficient condition. In addition, real solutions of the stationary
phase condition must exist in order to have a constructive interference
between the electrons 
across the system. The otherwise destructive interference damps the
oscillation not contributing to the coupling. Finally, if such behaviour
can be found even within the single-band model, for which the FS is
simple and has a single sheet, we may expect the occurrence of non-RKKY
periods in systems with more elaborate FS. Surfaces with more than one
sheet, such as non-magnetic transition metals for instance, are possible
candidates for presenting such periods. In those cases, the
interpretation of the results in terms of the RKKY theory may be
misleading.  

We are grateful to EPSRC and Royal Society of UK, and CNPq of Brazil for
financial support. We also would like to acknowledge useful discussions
with Dr. R. B. Muniz and Dr. A. Umerski.

\appendix
\section{}
\label{fourier}

Alternatively, within the single-band model, the Fourier coefficients
$C_{m_1,m_2}$ in section \ref{fcc110} can be calculated in the
following way. Bearing in mind the equivalence between 
$g_{0 n} = g_{n 0}$, Eq.(\ref{G}) can be concisely written as
\begin{equation}
G^{\uparrow}_{n 0} = {\tau_{\uparrow}\,
g_{0 n}\,(1-\tau_{\uparrow}\,g_{0 0}) \over V_{ex}}
\sum_{j=0}^\infty (\tau_{\uparrow}\, g_{n 0})^{2j} \,\,,
\label{GonA}
\end{equation}
and analogously,
\begin{equation}
G^{\downarrow}_{0 n} = - {\tau_{\downarrow}\,
g_{0 n}\,(1-\tau_{\downarrow}\,g_{0 0}) \over V_{ex}}
\sum_{j^\prime=0}^\infty (\tau_{\downarrow}\, g_{n 0})^{2j^\prime} \,\,.  
\label{GnoA}
\end{equation}
The product of Eqs.(\ref{GonA}) and (\ref{GnoA}) is 
\begin{equation}
G^{\uparrow}_{n 0}\,G^{\downarrow}_{0 n} = -
{\tau_{\uparrow}\tau_{\downarrow}\,(1-\tau_{\uparrow}\,g_{0
0})(1-\tau_{\downarrow}\,g_{0 0})  \over (V_{ex})^2} \sum_{j,j^\prime} 
(\tau_{\uparrow})^{2j}(\tau_{\downarrow})^{2j^\prime}\, (g_{n
0})^{2(j+j^\prime+1)} \,\,.
\label{prodA} 
\end{equation}
Writing the logarithmic function as a series, we have from
Eq.(\ref{Fwq})
\begin{equation}
{\cal F} = Im \left\{ \sum_{\ell=1}^\infty {(-1)^{\ell+1} \over
\ell}\left[\,{\cal T} \sum_{j,j^\prime}
(\tau_{\uparrow})^{2j}(\tau_{\downarrow})^{2j^\prime}\, (g_{n
0}^2)^{(j+j^\prime+1)}\right]^\ell \right\} \,\,,
\label{Fprime}
\end{equation}
where ${\cal T} = - 4
\tau_{\uparrow}\tau_{\downarrow}\,(1-\tau_{\uparrow}\,g_{0
0})(1-\tau_{\downarrow}\,g_{0 0})$. We recall that the summations above
have indices with different limits. Whereas the external summation 
index $\ell$ runs over the integers starting from $\ell = 1$ onwards,
the inner indices $j$ and $j^\prime$ have zero as their initial values.
Eq.(\ref{Fprime}) can be manipulated and rewritten as
\begin{equation}
{\cal F} = Im \left\{ \sum_{\ell=1}^\infty {(-1)^{\ell+1} \over \ell}
{\cal T}^\ell \sum_{j_1,j_1^\prime}...\sum_{j_\ell,j_\ell^\prime} 
(\tau_{\uparrow})^{2(j_1+...+j_\ell)}(\tau_{\downarrow})^{2(j_1^\prime+...
+j_\ell^\prime)}\,(g_{n
0}^2)^{(j_1+...+j_\ell+j_1^\prime+...+j_\ell^\prime+\ell)}\right\} \,\,.
\label{fprimeA}
\end{equation}
For a fixed $\ell$ there are twice as many inner summations, running
from $j_1$ and $j^\prime_1$ as far as $j_\ell$ and $j^\prime_\ell$. We 
define the integer $p$ as the exponent of the function $g^2_{n 0}$ in
Eq.(\ref{fprimeA}), i.e., $p =
j_1+j_1^\prime+...+j_\ell+j_\ell^\prime$. Note that $p$ runs over the
positive integers greater than zero.  

Squaring $g_{n 0}$ in Eq.(\ref{g0n2}) and inserting it into
Eq.(\ref{fprimeA}) we obtain the function ${\cal F}$ in terms of a
Fourier expansion. Clearly, the exponential $e^{i ( m_1 q_1 + m_2 q_2 )
n d}$ arises from terms of $(g_{n 0})^{2 p}$ such that $p = {m_1 + m_2
\over 2}$. The number of terms in turn depends on the number of 
possible combinations between the implicit indices $j$, $j^\prime$ and
$\ell$, and the larger $p$ is, the more combinations are possible.  

For $p = 1$ for instance, the only possible combination of the indices
is $j_1 = j_1^\prime = 0$ and $\ell = 1$. The terms
coming from this contribution correspond to the RKKY-limit expression in
Eq.(\ref{Frkky}), which indicates ${\cal F}_{RKKY}$ as a subset of the full
expression ${\cal F}$ and confirms the RKKY theory as a limit of the more
general approach used here \cite{paper1}. 

For $p = 2$, more combinations
are possible. In addition to $\ell = 2$ and $j_1 = j_1^\prime = j_2 =
j_2^\prime = 0$, four other combinations come from $\ell = 1$ and $j_1 +
j_1^\prime + j_2 + j_2^\prime = 1$. Note that the lowest order
coefficients are easily calculated through this alternative method but
as the exponent $p$ is increased, more terms arise and increase the
difficulty. Nevertheless, because we are searching for periods other
than the RKKY ones, the lowest order terms beyond the RKKY-limit are
expected to be the important corrections to the coupling. 

Finally, from the discussion above, two properties of the Fourier
coefficients are straightforwardly obtained. They are
$C_{-m_1,-m_2} = C_{m_1,m_2}^*$ and $C_{m_1,m_2} = 0$ unless $m_1 + m_2$
is even. These properties are also valid in the general case of section
\ref{general}.

\pagebreak
\pagebreak
\pagebreak
\pagebreak
\pagebreak
\pagebreak

\end{document}